\documentclass[english]{eptcs}
\usepackage[T1]{fontenc}
\usepackage[latin9]{inputenc}
\usepackage{calc}
\usepackage{amsmath,amssymb,amsthm}
\usepackage{stmaryrd}

\newtheorem{definition}{Definition}

\newtheorem{lemma}{Lemma}

\newcommand{ \DO}{{\bf do\ }}
\newcommand{ \OD}{{\bf od\ }}

\newcommand{ \bwhen}{{\bf when\ }}
\newcommand{ \bthen}{{\bf then\ }}

\newcommand{ \bend}{{\bf end}}

\newcommand{\e}{\mathsf{e}}

\newcommand{\defs}{~\hat{=}~}
\newcommand{\g}{{\sf g}}

\newcommand{\sched}{{\sf sched}}

\newcommand{\magic}{{\sf magic}}
\newcommand{\abort}{{\sf abort}}
\newcommand{\skipc}{{\sf skip}}
\newcommand{\true}{{\sf true}}
\newcommand{\false}{{\sf false}}

\makeatletter

\providecommand{\tabularnewline}{\\}

\makeatother

\usepackage{babel}

\begin{document}

\title{Concurrent Scheduling of Event-B Models
  \thanks{This research was supported by the EU funded FP7 project DEPLOY (214158). http://www.deploy-project.eu}
}
\def\titlerunning{Concurrent Scheduling of Event-B Models}

\author{Pontus Bostr\"om, Fredrik Degerlund, Kaisa Sere and Marina Wald\'en
\institute{\AA bo Akademi University, Dept. of Information Technologies\\
Turku Centre for Computer Science (TUCS)\\
Joukahainengatan 3-5, FIN-20520 \AA bo, Finland}
\email{\{pontus.bostrom, fredrik.degerlund, kaisa.sere, marina.walden\}@abo.fi}
}
\def\authorrunning{P. Bostr\"om, F. Degerlund, K. Sere and M. Wald\'en}
\maketitle
\begin{abstract}
Event-B is a refinement-based formal method that has been shown to be useful
in developing concurrent and distributed programs. Large models can be decomposed
into sub-models that can be refined semi-independently and executed in parallel. In this paper,
we show how to introduce explicit control flow for the concurrent sub-models in the form of event
schedules. We explore how schedules can be designed so that their application results in a
correctness-preserving refinement step. For practical application, two patterns for schedule
introduction are provided, together with their associated proof obligations. We demonstrate our
method by applying it on the dining philosophers problem.
\end{abstract}

\section{Introduction}

Event-B \cite{Abrial-10,rodin} is a state-based modelling framework with its roots in the guarded command
language and the Action Systems formalism \cite{BaKu83,BaSe91}. It advocates proof-based correct-by-construction design, abstraction, stepwise refinement and model decomposition as its main development strategies. 

In an Event-B model, events are chosen non-deterministically for
execution following the interleaving principle and assuming atomicity of events. Much of the effort in the refinement approach, especially down in the refinement chain, is about the modeller aiming at diminishing the non-determinism in the model and introducing more deterministic ways of choosing events for execution. In an extreme case we can think of the modeller encoding this by using explicit program counters in the events. Work on introducing more deterministic \emph{schedules} of events to Event-B has been studied extensively recently \cite{Bostrom-10,Hallerstede-10,Ilisov-09,Schneider-Treharne-Wehrheim-10}. The goal has been to avoid explicitly coding this scheduling information into the events. We base our approach on \cite{Bostrom-10}, which concerns sequential systems, and extend it to concurrent programs.

When models become large, decomposition strategies are used to focus on specific parts of the model. To be practical, such strategies need to support compositional
verification in the sense that the modeller can locally reason about properties of a decomposed part of the model even though the underlying Event-B assumption is that events are chosen for execution from the entire set of events in the model. Relying on the atomicity requirement for events and the interleaving semantics for Event-B models the distinct parts can be interpreted as concurrently executing models \cite{Hoang-Abrial-10}. We show here how
the scheduling approach of Bostr\"om \cite{Bostrom-10} can be extended so that we can apply it in a compositional manner focusing only on part(s), or sub-model(s), of the model. We turn these sub-models into tasks, giving each of them a schedule of its own. The main addition to the original approach for sequential programs is to handle the possible interferences the concurrently executing tasks might exhibit. This can also be seen as an extension, with explicit schedules, of the Hoang-Abrial approach \cite{Hoang-Abrial-10} to development of concurrent programs.

To facilitate practical use of our method, the schedules are introduced stepwise into a model via patterns. The patterns have associated
proof obligations needed for ensuring the correctness of the refinement step. As a result of the schedules,
the scheduling information contained in events can be expressed explicitly in the schedules. 

In this paper, we focus on developing concurrent programs following the stepwise refinement approach. Apart from the introduction of explicit schedules, concurrent programs are modelled within Event-B in a normal manner \cite{Abrial-10,Hoang-Abrial-10}. While Event-B models can be executed as such using a non-deterministic scheduler (``animation''), our approach is designed to be close to traditional programming languages and results in models that are more efficient to execute on a computer, since more control flow information is explicitly stated in the schedule than using only Event-B \cite{Bostrom-10}. The approach can also be used to replace parts of event behaviour with scheduling information as the scheduling concept as such is more general than what the focus is here.  The schedules actually give a process-oriented specification style for Event-B modeller complementing its state-based style \cite{Butler-00,Plosila-Sere-Walden-05}.

The rest of this paper is structured as follows. In section \ref{sec:Foundations},
we present the foundations needed to understand our approach. We discuss
set transformers (predicate transformers), the Event-B formalism and
model decomposition. In section \ref{sec:Dining-philosophers}, we introduce
a dining philosophers \cite{Hoare-85} Event-B model, which serves as a running
example. Section \ref{sec:Scheduling} presents our main contributions. We introduce
a scheduling language, show how schedules and tasks can be introduced, and demonstrate
how it is possible to tackle the problem of interference from interleaving tasks. In
section \ref{sec:Dining-philosophers-2}, we show how our framework
can be applied on the dining philosophers example model. Finally,
we sum the paper up in section \ref{sec:Conclusions-and-related},
where we also discuss related work and future perspectives.

\section{Foundations\label{sec:Foundations}}

\subsection{Event-B}

Event-B \cite{Abrial-10,Hallerstede-08} is a state-based modelling language.
Models in Event-B consist of a dynamic and a static part, referred to as \emph{machines}
and \emph{contexts}, respectively. The most important parts of a machine
are \emph{variables}, an \emph{invariant} and \emph{events}.
Contexts contain parts such as \emph{constants}, which can be referred
to from machines. The state space is made up of the variables $v_{1}$, ..., $v_{n}$
of types $\Sigma_{1}$, ..., $\Sigma_{n}$, and can be modelled as
the cartesian product $\Sigma=\Sigma_{1}\times...\times\Sigma_{n}$.
The events $E_1$, ..., $E_m$ modify the state space, and can be written
in the following general form \cite{Hallerstede-08}, where $k\in1..m$:
\begin{equation}
E_k \defs \bwhen G_k(v,c) ~\bthen v:|A_k(v,v',c) ~\bend.
\end{equation}
Here, $v$ represents the variables, $c$ the constants
seen by the machine, and the \emph{action} $v:|A_k(v,v',c)$ is the
nondeterministic assignment assigning $v$ any such values $v'$ for
which $A_k(v,v',c)$ holds. $G_k(v,c)$ represents the \emph{guard}, which is
a condition that must hold in order for the action to take place.
An event is said to be \emph{enabled} when its guard holds.
Each machine also contains a special event $\mathit{Initialisation}$ $\defs v:|A_0(v',c)$
that initialises the state space. Unlike other events, it is unguarded and does not
depend on a previous state. Events can be classified as \emph{ordinary}, \emph{convergent} or
\emph{anticipated}. This will be further explained in section
\ref{sub:Behavioural-semantics}. The invariant $I(v,c)$ is a predicate constraining
the values of the variables.

%
%
%
%

\subsection{Set transformers}


The events in Event-B can be viewed as set transformers \cite{Hallerstede-08}.
Our presentation of events as set transformers is similar to the presentation in \cite{Hallerstede-08}.

Consider a state space $\Sigma$. A set transformer is a function
$\mathcal{P}(\Sigma)\rightarrow\mathcal{P}(\Sigma)$ that tranforms
a set of states into another set of states. A weakest precondition set
transformer $S$ applied to a set $q$ returns the largest set $p$ from
which $S$ is guaranteed to reach a state in $q$.

We have the following definitions to give a set transformer semantics to
Event-B models:
\begin{equation}\label{eqn:event-b-set-def}
\begin{array}{lcl}
\Sigma & =  & \{v|\top\}\\
i & = & \{v|I(v,c)\}\\
g_k & = & \{v|G_k(v,c)\}\\
a_k & = & \{v\mapsto v'|A_k(v,v',c)\}\\
a_0 & = & \{v'|A_0(v',c)\}
\end{array}
\end{equation}
The set $i$ describes the subset of the state space where the invariant $I$ holds. Similarly, the sets $g_k$ ($k\in1..m$) represent the state space subsets where guard $G_k$ of the respective event $E_k$ is true. The relation $a_k$ describes the possible before-after states that can be achieved by the assignment of the respective event. Note that the initialisation results in a set $a_0$ instead of a relation, since it does not depend on the previous values of the variables. In this paper, we do not consider properties of constants $c$ separately, as it is not important at this level of reasoning. The axioms that describe the properties of the constants are here considered to be part of the invariant.

Let $g$ and $q$ be subsets of $\Sigma$, and $a$ be a relation. Furthermore, $S$, $S_{1}$ and
$S_{2}$ are arbitrary set transformers. The variables of $\Sigma$ are denoted $v$. We have
the following set transformers:
\noindent{}\begin{align}
&[a](q) \defs \{v|a[\{v\}]\subseteq q\} & \mathrm{(Nondeterministic~update)}\\
&\left[g\right](q) \defs \neg g\cup q & \mathrm{(Assumption)}\\
&\{g\}(q) \defs g\cap q & \mathrm{(Assertion)}\\
&(S_{1}\sqcap S_{2})(q) \defs S_{1}(q)\cap S_{2}(q) & \mathrm{(Nondeterministic~choice)}\\
&S_{1};\, S_{2}(q) \defs S_{1}(S_{2}(q)) & \mathrm{(Sequential~composition)}\\
&S^{\omega}(q) \defs \mu X.(S;\, X\sqcap \skipc{})(q) & \mathrm{(Strong~iteration)}\\
&S^{*}(q) \defs \nu X.(S;\, X\sqcap \skipc{})(q) & \mathrm{(Weak~iteration)}\\
&\skipc(q) \defs q & \mathrm{(Stuttering)}\\
&\magic(q) \defs \true & \mathrm{(Miracle)}\\
&\abort(q) \defs \false & \mathrm{(Aborting)}
\end{align}
\noindent{}Here, $\true$ and $\false$ are notations representing the sets
$\Sigma$ and $\emptyset$, respectively. This is because of convenience
as well as the fact that the same notation is used in weakest precondition
predicate transformers. We will also in general use predicate notation
for describing subsets of the state space. (Nondeterministic) update
is used to assign values to variables in the state space,
of which the stuttering set transformer $\skipc{}$ is a special case, which leaves
the state unmodified. The set transformer $\magic$ achieves the desired postcondition
(even $\false$) from any state, whereas $\abort$ does not guarantee
to achieve any postcondition $q$ from any state. Not even termination
is guaranteed. Assumption and assertion both behave as $\skipc{}$ when $g$
is true, but when false, assumption behaves as $\magic$, whereas assertion
behaves as $\abort$. Nondeterministic choice represents demonic choice between
set transformers,
and sequential composition combines set transformers in a sequential manner.
An important property of demonic choice is that miraculous behaviour
is avoided whenever possible, whereas aborting behaviour is always preferred.
This is demonstrated by the following theorems, which follow directly
from the definitions:
\begin{equation}
\begin{array}{ll}
\magic\sqcap S = S\\
\abort\sqcap S = \abort\\
\end{array}\\
\end{equation}
The following properties can easily be derived, and the proofs can
also be found in \cite{Back-vonWright98}:
\begin{equation}
\begin{array}{lll}
\magic;\, S = \magic & ~~~~~~~ & \abort;\, S = \abort \\
\{g\};\,[h] = \{g\} & ~ & [g];\,\{h\} = [g] \\
\{g\cap h\} = \{g\};\,\{h\} & ~ & [g\cap h] = [g];\,[h]\\
\end{array}\\
\\
\end{equation}
The iteration set transformers are used to achieve repeated execution. Iteration
has been thoroughly discussed by Back and von Wright
\cite{Back-vonWright98, Back-vonWright-99}, and is
only shortly summarised here. In both strong and weak iteration ($S^{\omega}$
and $S^{*}$, respectively), the set transformer $S$ is repeatedly executed
a demonically chosen number of times. In strong iteration, the number
of executions may be infinite, whereas for weak iteration it is guaranteed
to be finite. Important theorems regarding iteration include the following
\emph{unfolding} rules:
\begin{equation}\label{eqn:rule-unfolding}
\begin{array}{l}
S^{\omega}=S;\, S^{\omega}\sqcap \skipc\\
S^{*}=S;\, S^{*}\sqcap \skipc\\
\end{array}
\end{equation}

The set of states in which a set transformer $S$ does not
behave miraculously is called the guard of $S$. The guard $\g(S)$ is given
as:
\begin{equation}
\g(S) \defs \neg S(\false)
\end{equation}

We can now interpret an event $E_k$ from (1) as a set transformer. Using the
definitions from (2), we can now give the set transformer $[E_k]$ for $E_k$ as
\cite{Hallerstede-08}:
\begin{equation}
[E_k] \defs [g_k]; [a_k]
\end{equation}
For a set of events, $\{E_1,\ldots , E_m\}$, we will use the denotion $[E]$
for the expression $[E_1]\sqcap \ldots\sqcap[E_m]$.

\subsection{Refinement}
Refinement is an important concept in Event-B. In this paper, we are mainly interested
in refinement on the set transformer level, where it can be defined as \cite{Back-vonWright98}:
\begin{equation}
 S_1\sqsubseteq S_2 \defs \forall s.S_1(s)\subseteq S_2(s)
\end{equation}
\noindent{}Here, $S_1$ and $S_2$ are set transformers. The intuitive interpretation
of $S_1\sqsubseteq S_2$ is that if $S_1$ will reach a state in a set $s$,
then so will $S_2$. We say that $S_1$ and $S_2$ are (refinement) equivalent
if and only if $S_1\sqsubseteq S_2$ and $S_2\sqsubseteq S_1$. The relation between
the set transformer view of refinement and a proof obligations approach
has been studied in \cite{Hallerstede-08}.

A set transformed $S$ is said to behave miraculously when executed in a state in the set $S(\false)$, i.e.
when the execution of $S$ results in a post-state belonging to the empty set. We typically want to avoid
introduction of more miraculous behaviour during refinement. Given a set transformer $S_1$ and a refinement $S_2$, $S_2$
does not exhibit more miraculous behaviour than $S_1$ if $S_1(\false) = S_2(\false)$.

\subsection{Behavioural semantics}\label{sub:Behavioural-semantics}
We aim at using Event-B for construction of concurrent programs. Ultimately we like to show that a (concurrent) program $S$ is correct given a precondition $P$ and a postcondition $Q$. This correctness requirement is expressed in the Hoare triple:
\begin{equation}
\{P\}~S~\{Q\}
\end{equation}
As the basis for our method, we use the development method for concurrent programs in \cite{Hoang-Abrial-10}. In this approach, the concurrent programs are built from atomic events in the same way as sequential programs are constructed  \cite{Abrial-10}. The program $S$ is considered to consist of a collection of events. Note that there is no control flow other than non-deterministic choice of enabled events. Using the refinement based approach of Event-B,  the program $S$ that satisfies the pre/post-specification is derived stepwise. In order to use the refinement process to develop programs, the pre-/post-specification first has to be encoded into an initial Event-B model. This model has a specific structure \cite{Abrial-10}: it  has an initialisation event $\mathit{init}$, progress events $\mathit{prog}$ and a finalisation event $\mathit{fin}$. The events $\mathit{prog}$ model (non-deterministically) the computation of the program, while $\mathit{fin}$ models the post-condition $Q$ as a guard. The precondition is encoded in an external context machine. The semantics of an Event-B model $M$ specifying a sequential program is in this setting:
\begin{equation}
\label{eqn:Beh-semantics}
M\defs\left[\mathit{init}\right];[\mathit{prog}]^{*};[\mathit{fin}]
\end{equation}
The system is first initialised, then $\mathit{prog}$ is executed until the postcondition given by $\mathit{fin}$ becomes true. The program can then terminate. The progress events $\mathit{prog}$ are later refined to create a deterministic algorithm to reach the postcondition. We will also later need to show that the refinements $E$ of $\mathit{prog}$ terminate \cite{Abrial-10}, i.e. $[E]^\omega=[E]^{*}$, as we are interested in total correctness. We assume that all Event-B models in the rest of the paper have this structure. Each event should maintain the invariant and therefore we assume that there is an invariant assertion $\{i\}$ implicitly given before and after each event.


We previously mentioned that events can be classified as \emph{ordinary}, \emph{convergent}
or \emph{anticipated}. This is relevant from a behavioural semantics point of view. Events are normally
classified as ordinary, but it is sometimes necessary to prove that execution of events from a group will
eventually terminate. All events belonging to this group should then be labelled as convergent. In practice,
the termination property is proven by introducing a variant, and by showing that it is decreased
by all convergent events. There is also the possibility of classifying events as anticipated. Labelling
an event as anticipated indicates that it will be classified as convergent in a later refinement step, whereby
the proof is postponed until further down the refinement chain. The notions anticipated or convergent should be
for the events $\mathit{prog}$ to guarantee that the model eventually terminates.

\subsection{Decomposition\label{sub:Decomposition}}

In order for a refinement based development method to be scalable there should be a way to decompose specifications into smaller parts that can be independently developed. The verification of refinement should thus be compositional, i.e., refinement of the individual parts should yield a refinement of the whole system.

Here we will use a decomposition approach based on shared variables \cite{Abrial-10, Abrial-Hallerstede-07}. Following
this approach, a model can be decomposed into sub-models that can themselves be further decomposed. The set of
sub-models forms the complete system model.
\begin{definition}{Sub-model.}\label{def:module}
A sub-model is given as a 7-tuple $(v,x,E,X,I,init,\mathit{fin})$, where $v$ and $x$ are sets of variables, $E$ and $X$ are sets of events, $I$ the invariant, $init$ the initialisation and $\mathit{fin}$ the finalisation.
\end{definition}
\noindent{}The variables $v$ are only visible inside the sub-model, and will be referred to as internal variables. Variables $x$ are shared with other components and will be called external variables. The events $E$ can refer to both $v$ and $x$. Since they (also) manipulate the internal variables of the sub-model, they are denoted the internal events. The external events, $X$, are abstractions that only refer to the external variables $x$ modelling the effects of events of other components. Hence, each event in $X$ has a corresponding internal event in another component. The initialisation of a sub-model is given by event $init$ and the loop termination guard is given by event $\mathit{fin}$. Note that a traditional Event-B model can be seen as a sub-model where the sets of external events and external variables are empty. A sub-model $(v,x,E,X,I,init,fin)$ can be (further) decomposed into sub-models:
\begin{equation}
(v,x,E,X,I,init,fin)=(v_1,x_1,E_1,X_1,I_1,init_1,fin_1) \parallel (v_2,x_2,E_2,X_2,I_2,init_2,fin_2)\nonumber
\end{equation}
The parallel composition of the sub-models is defined as:
\begin{equation}
\begin{array}{ll}
~&(v_1,x_1,E_1,X_1,I_1,init_1,fin_1)\parallel(v_2,x_2,E_2,X_2,I_2,init_2,fin_2)\\
\defs&(v_1\cup v_2,(x_1\cup x_2)\backslash(v_1\cup v_2), E_1\cup E_2, (X_1\cup X_2)\backslash (E_1\cup E_2),I_1 \wedge I_2, init_1\parallel init_2,$ $ fin_1\parallel fin_2)
\end{array}
\end{equation}
The parallel composition of two events is given as:
\begin{equation}
\begin{array}{ll}
~&\bwhen G~ \bthen v:|S~ \bend \parallel \bwhen H~ \bthen w:|R~ \bend~\\
\defs&\bwhen G\wedge H~  \bthen v,w:|S\wedge R~  \bend
\end{array}
\end{equation}
The semantics $[M_1\parallel M_2]$ of a the parallel composition $M_1\parallel M_2$ is given as:
\begin{equation}
\left[M_1\parallel M_2 \right]\defs init_1\parallel init_2; ([E_1\cup E_2 \cup ((X_1\cup X_2)\backslash (E_1\cup E_2))] )^{*};[\neg \g(fin_1\parallel fin_2)]
\end{equation}
The composition can be extended to arbitrary many components by recursively merging components pairwise. Since we want to do compositional proofs of refinement, we need to show that refinement of the individual sub-models lead to refinement of the entire system. First we need to prove that the external events provide abstractions of their internal counterparts $\{i_1\cap i_2\};[X_1]\sqsubseteq [E_2]\sqcap [X_2]$ and $ \{i_1\cap i_2\};[X_2]\sqsubseteq [E_1]\sqcap [X_1]$. To compositionally prove the refinement $[M_1\parallel M_2]\sqsubseteq [M_1'\parallel M_2]$, we then only need to prove the refinement $[M_1]\sqsubseteq [M_1']$, see  \cite{BaWr03}.

We need to model that external events are executed a finite number of times, as they model the finite execution of their internal counterparts in other sub-models. Since these external events are not necessarily terminating by themselves, strong iteration cannot be used for describing behaviour of sub-models. The use of weak iteration can be seen as compositionally verifying partial correctness of a program, since termination is not ensured by set transformer refinement. However, we want to prove total correctness of the complete system. Since we in this approach \cite{Abrial-10,Hoang-Abrial-10} label the events $E$ as anticipated or convergent, we show that the model will eventually terminate. Hence, total correctness follows from partial correctness in combination with the Event-B proof obligations that ensure termination \cite{Back-vonWright98,Back-vonWright-99}.

\section{Dining philosophers case study\label{sec:Dining-philosophers}}

\subsection{Problem description}

We are now ready to introduce a model of the dining philosophers \cite{Hoare-85}, which will
serve as a running example. In this section, we show the initial model,
we refine it, as well as decompose it into sub-models. The dining
philosophers scenario can be described as follows. There are four philosophers
sitting around a round table. Each philosopher has a plate in front of him,
and there is a fork placed between each pair of adjacent plates. Each
philosopher always does one of two things: think and eat, but not both at
the same time. Furthermore, in order to eat, a philosopher must pick up
both of the two forks located next to his plate. A philosopher can also
drop a fork back into its original position, but only after he has eaten. 

The basic problem is that if the philosophers pick up the forks arbitrarily,
there may be deadlocks. For example, if each philosopher picks up
his right fork, there will not be any forks available anymore, and
no philosopher will have enough forks to eat. Since a philosopher
will not drop a fork until he has eaten, there will be a deadlock.
One well-known solution to this problem is to assign a number to each
fork, and enforce that each philosopher picks up the adjacent fork
with the lowest number first. In our case study we assume that we have
four philosophers and number the forks as follows:
Philosopher 1 can access forks 1 and 2, philosopher 2 accesses forks
2 and 3, philosopher 3 uses forks 3 and 4, while philosopher 4 has
access to forks 1 and 4.

\subsection{Modelling and refinement}

Initially we model the scenario as an abstract Event-B machine, where
the four philosophers eat in a non-deterministic order. We only model
one round, so each philosopher will only eat once. We introduce the
variables \emph{ph1eaten} thru \emph{ph4eaten}, to model whether each
philosopher has eaten. The event \emph{Intialisation} sets these variables
to FALSE. The events \emph{Ph1Eat} thru \emph{Ph4Eat} for the four philosophers
then represent the progress of the model. They model that a philosopher eats
which has not yet eaten by setting the corresponding variable to TRUE. Finally,
event \emph{Finalisation} checks that all four philosophers have eaten.
The \emph{Initialisation} and \emph{Finalisation} events are classified
as ordinary events, whereas \emph{Ph1Eat}, ..., \emph{Ph4Eat} are convergent,
since they correspond to the $\mathit{prog}$ variables in (\ref{eqn:Beh-semantics}).
We now have:\\

{\small
\begin{center}
\begin{tabular}{ccc}
\framebox{\begin{minipage}[t]{0.2\columnwidth}%
\textbf{variables}

\hspace{5mm}\emph{ph1eaten}

\hspace{5mm}\emph{ph2eaten}

\hspace{5mm}\emph{ph3eaten}

\hspace{5mm}\emph{ph4eaten}%
\end{minipage}} & %
\framebox{\begin{minipage}[t]{0.3\columnwidth}%
\textbf{invariant}

\hspace{5mm}\emph{ph1eaten} $\in$ BOOL

\hspace{5mm}\emph{ph2eaten} $\in$ BOOL

\hspace{5mm}\emph{ph3eaten} $\in$ BOOL

\hspace{5mm}\emph{ph4eaten} $\in$ BOOL %
\end{minipage}}& %
\framebox{\begin{minipage}[t]{0.3\columnwidth}%
Initialisation (\emph{ordinary}) $\defs$

\textbf{begin}

\hspace{5mm}\emph{ph1eaten} := FALSE

\hspace{5mm}\emph{ph2eaten} := FALSE

\hspace{5mm}\emph{ph3eaten} := FALSE

\hspace{5mm}\emph{ph4eaten} := FALSE 

\textbf{end}%
\end{minipage}}\tabularnewline
\end{tabular}
\par\end{center}

\begin{center}
\begin{tabular}{cc}
\framebox{\begin{minipage}[t]{0.35\columnwidth}%
Ph1Eat (\emph{convergent}) $\defs$

\textbf{when}

\hspace{5mm}\emph{ph1eaten} = FALSE

\textbf{then}

\hspace{5mm}\emph{ph1eaten} := TRUE

\textbf{end}%
\end{minipage}} & %
\framebox{\begin{minipage}[t]{0.3\columnwidth}%
Finalisation (\emph{ordinary}) $\defs$

\textbf{when}

\hspace{5mm}\emph{ph1eaten} = TRUE

\hspace{5mm}\emph{ph2eaten} = TRUE

\hspace{5mm}\emph{ph3eaten} = TRUE

\hspace{5mm}\emph{ph4eaten} = TRUE

\textbf{then}

\hspace{5mm}\emph{skip}

\textbf{end}%
\end{minipage}}\tabularnewline
\end{tabular}\\

\par\end{center}
}

In the first refinement step we
introduce the forks, which are modelled as variables \emph{fork1}
thru \emph{fork4}. They are of type 0..4 to represent which philosopher
that currently holds the fork. Value 0 represents the fork lying on
the table. All forks are initialised to this value. There are 16
new events in this refinement step: two for each of the four philosophers
getting their adjacent forks (e.g. \emph{Ph3GetFork3} and \emph{Ph3GetFork4}), and
two events for each philosopher releasing the corresponding forks (e.g. \emph{Ph3RelFork4} and \emph{Ph3RelFork3}). Note that philosopher 4 uses forks 1 and 4. 

In order to be able to prove that the new events will not take over the execution, we classify them as convergent and give a variant that they decrease. There is no variable that can be used as a variant, but when each new event is executed it will disable itself and it will not be enabled again. Hence, we define a function $v$ as follows:

\[
{\small
\begin{array}{ll}
v = \{&(FALSE,FALSE,FALSE) \mapsto 5, \\
~& (TRUE,FALSE,FALSE) \mapsto 4, \\
~& (TRUE,TRUE,FALSE) \mapsto 3, \\
~& (TRUE,TRUE,TRUE) \mapsto 2, \\
~& (TRUE,FALSE,TRUE) \mapsto 1, \\
~& (FALSE,FALSE,TRUE) \mapsto 0 \}
\end{array}
}
\]

\noindent{}The first and second dimension of the triple correspond to whether a philosopher is holding his left or right fork, respectively. The third one indicates whether he has already eaten or not. The variant is then formed as a sum of the values of function $v$ applied on the variables of each philosopher. The refined model is now as follows:\\

{\small
\begin{center}
\begin{tabular}{ccc}
\framebox{\begin{minipage}[t]{0.13\columnwidth}%
\textbf{variables}

\hspace{5mm}\emph{fork1}

\hspace{5mm}\emph{fork2}

\hspace{5mm}\emph{fork3}

\hspace{5mm}\emph{fork4}

\hspace{5mm}\emph{ph1eaten}

\hspace{5mm}\emph{ph2eaten}

\hspace{5mm}\emph{ph3eaten}

\hspace{5mm}\emph{ph4eaten}%
\end{minipage}} & %
\framebox{\begin{minipage}[t]{0.5\columnwidth}%
\textbf{invariant}

\hspace{5mm}\emph{fork1} $\in$ 0..4

\hspace{5mm}\emph{fork2} $\in$ 0..4 

\hspace{5mm}\emph{fork3} $\in$ 0..4 

\hspace{5mm}\emph{fork4} $\in$ 0..4 

\hspace{5mm}\ldots

\textbf{variant}

\hspace{5mm}$v(bool(fork1=1),bool(fork2=1),ph1eaten)$

\hspace{10mm}$+ v(bool(fork2=2),bool(fork3=2),ph2eaten)$

\hspace{10mm}$+ v(bool(fork3=3),bool(fork4=3),ph3eaten)$

\hspace{10mm}$+ v(bool(fork1=4),bool(fork4=4),ph4eaten)$
\end{minipage}}& %
\framebox{\begin{minipage}[t]{0.25\columnwidth}%
Initialisation (\emph{ordinary}) $\defs$

\textbf{begin} 

\hspace{5mm}\emph{fork1} := 0

\hspace{5mm}\emph{fork2} := 0

\hspace{5mm}\emph{fork3} := 0

\hspace{5mm}\emph{fork4} := 0

\hspace{5mm}\emph{ph1eaten} := FALSE

\hspace{5mm}\emph{ph2eaten} := FALSE

\hspace{5mm}\emph{ph3eaten} := FALSE

\hspace{5mm}\emph{ph4eaten} := FALSE

\textbf{end} %
\end{minipage}} \tabularnewline
\end{tabular}
\par\end{center}

\begin{center}
\begin{tabular}{ccc}
\framebox{\begin{minipage}[t]{0.28\columnwidth}%
Ph1GetFork1 (\emph{convergent}) $\defs$

\textbf{when} 

\hspace{5mm}\emph{fork1} = 0

\hspace{5mm}\emph{ph1eaten} = FALSE

\textbf{then} 

\hspace{5mm}\emph{fork1} := 1

\textbf{end} %
\end{minipage}} & %
\framebox{\begin{minipage}[t]{0.28\columnwidth}%
Ph1GetFork2 (\emph{convergent}) $\defs$

\textbf{when} 

\hspace{5mm}\emph{fork1} = 1

\hspace{5mm}\emph{fork2} = 0

\hspace{5mm}\emph{ph1eaten} = FALSE

\textbf{then} 

\hspace{5mm}\emph{fork2} := 1

\textbf{end} %
\end{minipage}} & %
\framebox{\begin{minipage}[t]{0.28\columnwidth}%
Ph1Eat (\emph{convergent}) $\defs$

\textbf{when} 

\hspace{5mm}\emph{fork1} = 1

\hspace{5mm}\emph{fork2} = 1

\hspace{5mm}\emph{ph1eaten} = FALSE

\textbf{then} 

\hspace{5mm}\emph{ph1eaten} := TRUE

\textbf{end}%
\end{minipage}} \tabularnewline
\end{tabular}
\par\end{center}

\begin{center}
\begin{tabular}{ccc}
\framebox{\begin{minipage}[t]{0.28\columnwidth}%
Ph1RelFork2 (\emph{convergent}) $\defs$

\textbf{when} 

\hspace{5mm}\emph{fork2} = 1

\hspace{5mm}\emph{ph1eaten} = TRUE

\textbf{then} 

\hspace{5mm}\emph{fork2} := 0

\textbf{end} %
\end{minipage}} & %
\framebox{\begin{minipage}[t]{0.28\columnwidth}%
Ph1RelFork1 (\emph{convergent}) $\defs$

\textbf{when} 

\hspace{5mm}\emph{fork2}= 0

\hspace{5mm}\emph{fork1}= 1

\hspace{5mm}\emph{ph1eaten} = TRUE

\textbf{then} 

\hspace{5mm}\emph{fork1} := 0

\textbf{end} %
\end{minipage}} & %
\framebox{\begin{minipage}[t]{0.28\columnwidth}%
Finalisation (\emph{ordinary}) $\defs$

\textbf{when} 

\hspace{5mm}\emph{fork1} = 0

\hspace{5mm}\emph{fork2} = 0

\hspace{5mm}\emph{fork3} = 0

\hspace{5mm}\emph{fork4} = 0

\hspace{5mm}\emph{ph1eaten} = TRUE

\hspace{5mm}\emph{ph2eaten} = TRUE

\hspace{5mm}\emph{ph3eaten} = TRUE

\hspace{5mm}\emph{ph4eaten} = TRUE

\textbf{then} 

\hspace{5mm}\emph{skip}

\textbf{end} %
\end{minipage}}\tabularnewline
\end{tabular}\\

\par\end{center}
}

\noindent{}Note that when the $v$ function is called, the fork variables are not
directly passed as parameters. Instead, we check whether the currently evaluated
philosopher holds the fork or not. The \emph{bool} function is a technicality of
Event-B that is needed to convert the result of the comparison into a value of
BOOL.

The events corresponding to philosophers 2, 3 and 4 eating, as well as
picking up and releasing their respective forks are analogous to the
events of philosopher 1, and are thus not shown here. 
We now have a refined model for the four philosophers eating,
and in the next subsection we will decompose this model.

\subsection{Decomposition}

In the decomposition step we separate the functionality of the four philosophers
in such a way that each philosopher constitutes a sub-model of its own. The partitioning
we achieve is shown in the table below. Since philosophers 2 and 4 share fork 2 and
fork 1, respectively, with philosopher 1, the external events of sub-model 1 are
Ph2GetFork2, Ph2RelFork2, Ph4GetFork1 and Ph4RelFork1. Analogous reasoning is used
to find the external events of the other sub-models.\\

\begin{tabular}{|c||c|c|c|c|}
\hline
 & Sub-model 1 & Sub-model 2 & Sub-model 3 & Sub-model 4\tabularnewline
\hline
\hline
Internal & Ph1Eat & Ph2Eat & Ph3Eat & Ph4Eat\tabularnewline

events & \hspace{2mm} Ph1GetFork1 \hspace{2mm}  & \hspace{2mm} Ph2GetFork2 \hspace{2mm}  & \hspace{2mm} Ph3GetFork3 \hspace{2mm} & \hspace{2mm} Ph4GetFork1 \hspace{2mm} \tabularnewline

 & Ph1RelFork1 & Ph2RelFork2 & Ph3RelFork3 & Ph4RelFork1\tabularnewline

 & Ph1GetFork2 & Ph2GetFork3 & Ph3GetFork4 & Ph4GetFork4\tabularnewline

 & Ph1RelFork2 & Ph2RelFork3 & Ph3RelFork4 & Ph4RelFork4
\tabularnewline
\hline  
External & Ph2GetFork2 & Ph1GetFork2 & Ph2GetFork3 & Ph1GetFork1\tabularnewline

events & Ph2RelFork2 & Ph1RelFork2 & Ph2RelFork3 & Ph1RelFork1\tabularnewline

 & Ph4GetFork1 & Ph3GetFork3 & Ph4GetFork4 & Ph3GetFork4\tabularnewline

 & Ph4RelFork1 & Ph3RelFork3 & Ph4RelFork4 & Ph3RelFork4\tabularnewline
\hline
\end{tabular}\\

\section{Concurrent programs\label{sec:Scheduling}}


This far, we have considered model decomposition, resulting in sub-models that can be refined semi-independently. We are now ready to examine how these sub-models can be executed in a concurrent or parallel setting. This problem has been studied in \cite{Hoang-Abrial-10}, which is a case study showing how to decompose Event-B models into concurrently executing sub-models. Here we extend this approach by giving sub-models explicit flow control in the form of event schedules, instead of the traditional nondeterministic choice. An important concept in our approach is the concept of \emph{tasks}, which we define as follows:
\begin{definition}{Task.}
A task is an 8-tuple $(v,x,E,X,I,init,fin,S)$ where $v$ are the internal variables, $x$ the external variables, $E$ the internal events, $X$ the external events, $I$ the invariant, $init$ the initialisation, $fin$ the loop termination condition, and $S$ is a schedule conforming to the syntax in (\ref{eqn:scheduling-language}) concerning the internal events $E$. 
\end{definition}
\noindent{}Since all coordinates, except for $S$, are the same as in a sub-model, a task can be seen as an extension of the sub-model concept. Whereas the events of traditional decomposed sub-models are executed nondeterministically, the internal events of a task are scheduled according to $S$. The schedule $S$ may only consist of internal events, and the set of events in the schedule is denoted $\e(S)$. We assume that $E$ = $\e(S)$, since if an internal event was not included in the schedule, it would never be executed.

\subsection{Scheduling language}

In order to describe schedules of events we give a small scheduling language \cite{Bostrom-10}, which adheres to the following syntax:\\
\begin{equation}\label{eqn:scheduling-language}
\begin{array}{lcl}
S &::= &PS \rightarrow S~ |~ PS\\
PS& ::=& \DO S~ \OD~ |~ S_1 \talloblong\ldots\talloblong S_n~ |~ E~ |~ \{g\}
\end{array}
\end{equation}
 Here $\rightarrow$ represents sequential composition, $\talloblong$ non-deterministic choice, $\DO\OD$ is a loop, $E$ an event and $\{g\}$ is an assertion. 
 
 \subsection{Semantics of tasks}

The semantics of schedules is given using a function $\sched$ that maps each schedule to the corresponding set transformer as in \cite{Bostrom-10}. However, when scheduling the events in a task we need to consider interference from other tasks. A goal of the scheduling language is to be able to express schedules of internal events in such a way that interference from external events does not have to be explicitly taken into account. Such interference freedom is instead proven separately. We now recursively define a function $\sched(S,X)$ where $S$ is a schedule, $X$ is the set of external events. 
\begin{equation}
\begin{array}{lcl}
\sched(PS\rightarrow S,X) & = & \sched(PS,X);\sched(S,X)\\
\sched(\DO S~\OD,X) & = & ([\g([\e(S)\cup X])];\sched(S,X))^{*}; [\neg \g([\e(S)\cup X])]\\
\sched(S_1 \talloblong\ldots\talloblong S_n,X) &=& \sched(S_1,X)\sqcap\ldots\sqcap \sched(S_n,X)\\
\sched(E,X) & = & [X]^{*};[E];[X]^{*}\\
\sched(\{g\},X) & = & \{g\}
\end{array}
\end{equation}
The scheduling function takes the schedule $S$, as well as the set of external events $X$
as input and outputs a set transformer containing both internal and external
events.  An arbitrary (but finite) number of external events $X$ can occur before and after an internal event $E$ in a schedule. This is modelled by the set transformer $[X]^{*}$ on both sides of the event.

Consider a system consisting of two tasks $T_1=(v_1,x_1,E_1,X_{1},init_1,fin_1,S_1)$ and $T_2=(v_2,x_2,E_2,X_{2},$ $init_2,fin_2,S_2)$. To find the complete system behaviour, we need to compose the tasks, i.e. obtain $T_1\parallel T_2$. However, the number of interleavings of atomic set transformers grows exponentially with the length of the schedule \cite{deRoever-01}. Hence, we need an appropriate approach to reason about the interleavings in order to make refinement proofs manageable. Here we make the restriction that we only consider tasks where the set transformers obtained after scheduling can be decomposed into a loop containing the demonic choice of atomic set transformers. This is an extension of the approach used in \cite{Hoang-Abrial-10}, where the programs are built from atomic {\em events} that are chosen non-deterministically for execution. Composition of such tasks can be easily handled \cite{BaWr03}.  We have the following requirement for schedulability in our approach:
\begin{equation}\label{eqn:decompostion-equality}
\exists S_{11},\ldots,S_{1n}\cdot \sched(S_1,X_{1})=(S_{11}\sqcap\ldots \sqcap S_{1n}\sqcap [X_{1}])^{*};[fin_1]
\end{equation}
where all $S_{1i}$  are atomic compositions of internal events. Using these atomic set transformers we can now use the traditional parallel composition \cite{BaWr03}. The semantics of the composition of the whole system $T_1\parallel T_2$ is now given as:
\begin{equation}
\left[T_1\parallel T_2\right] \defs [init_1\parallel init_2]; ((\sqcap_i S_{1i}) \sqcap (\sqcap_j S_{2j}))^{*};[fin_1\parallel fin_2]
\end{equation}
This approach thus extends the decomposition method in \cite{Abrial-Hallerstede-07,Hoang-Abrial-10} with the possibility to reason about groups of sequentially scheduled events, instead of only individual ones. However, to find the groups $S_{11},\ldots,S_{1n}$ is in general non-trivial. Here we will give special cases encoded as {\em patterns} to make the verification of schedules manageable in practise.


\subsection{Introduction of schedules}
\label{sub:Introduction-of-schedules}

Schedules are introduced for the sub-models as a refinement step, in which we convert sub-models into tasks. The introduction of schedules has to constitute a refinement step in order to ensure that the properties we have already proved for the models before introduction of schedules are preserved. Note that we do not support scheduling of anticipated events, so
they have to be turned into convergent ones before the introduction of schedules.

We now need to show for the two tasks $T_1=(v_1,x_1,E_1,X_{1},init_1,fin_1,S_1)$ and $T_2=(v_2,x_2,E_2,X_{2},init_1,$ $fin_1,S_2)$:
\begin{equation}
[M_1\parallel  M_2] \sqsubseteq [T_1 \parallel T_2] 
\end{equation}
where sub-model $M_i$ corresponds to task $T_i$ as $M_i=(v_i,x_i,E_i,X_{i},init_i,fin_i)$. As in the traditional decomposition method, we can use external events to perform compositional proofs of refinement. Here we rely on the property (\ref{eqn:decompostion-equality}) to decompose schedule  $\sched(S_i,X_{i})$ into a loop consisting of atomic set transformers.  We need to show that  for all tasks $T_i$ \cite{BaWr03}:
\begin{align}
&\{i_1\cap i_2\};[X_{ij}] \sqsubseteq S_{kj}\label{eqn:external-event-scheduled}\\
&([\e(S_i)] \sqcap [X_{i}])^{*}; [fin_i]\sqsubseteq \sched(S_i,X_{i})\label{eqn:refinement-scheduled}
\end{align}
In (\ref{eqn:external-event-scheduled}) we assume that for any external event $X_{ij} \in X_i$, there is one corresponding atomic set transformer $S_{kj}$ in another task $T_k$. To give a practical approach to the decomposition of schedules required by (\ref{eqn:decompostion-equality}), we give patterns that give generic instantiations of the quantified variables.  In the patterns we rely on special cases of scheduling constructs where we know we can prove (\ref{eqn:external-event-scheduled}) and (\ref{eqn:refinement-scheduled}).  Patterns thus encode reusable schedule structures. One such case is when the introduction of sequential behaviour does not alter the behaviour of the sub-model. Another useful special case is when the introduction of sequential behaviour does not modify the externally visible behaviour of a sub-model. We use the same scheduling approach as in \cite{Bostrom-10}, where patterns are applied on schedules stepwise and we prove that each pattern application leads to a refinement of the previous application.

A pattern consists of a \emph{precondition}, a \emph{schedule}, a \emph{result} and a number of \emph{assumption}. The precondition predicate describes under which conditions the pattern is applicable. The schedule part describes what schedule the pattern is intended for, and the result part gives the set transformer that is produced when the pattern is applied. The assumptions are extra conditions that have be fulfilled in order to use the pattern.

\paragraph{Pattern 1}
The first pattern, $P_1$, introduces sequential behaviour into a sub-model.
\begin{equation}\label{eqn:scheduling-pattern-1}
\begin{array}{l}
 P_{1}(E_1,h,g,S, X)\defs \\
 \begin{array}{lcl}
 \mathrm{Precondition} &: & h\\
 \mathrm{Schedule} &: &  E_1\rightarrow \{g\}\rightarrow S  \\
 \mathrm{Result} &: & \{h\}; X^{*}; E_1; X^{*}; \{g\}; \sched(S,X) \\
 \mathrm{Assumption~ 1} &: &  h\subseteq \neg \g(\e(S))  \\
 \mathrm{Assumption~ 2} &: & g\subseteq \neg \g(E_1)\\
 \mathrm{Assumption~ 3} &: & \{g\};(X\sqcap \e(S))\sqsubseteq (X\sqcap \e(S));\{g\} \\
 \mathrm{Assumption~ 4} &: & \{h\};X\sqsubseteq X;\{h\} \\
 \mathrm{Assumption~5} & : & E_{1}=E_{1};\{g\}\\
\end{array}\\
\end{array}
\end{equation}
The first assumption states that the precondition $h$ implies that the events following $E_1$ are disabled. The second assumption states that $g$ ensures that $E_1$ is disabled. Context information cannot be propagated in schedules without taking interference into account. Hence we need assumptions 3 and 4 to state that $g$ and $h$ are invariant with respect to the environment. Furthermore, $g$ should also be invariant for all events in the schedule $S$. The last assumption states that $E_1$ will establish $g$. We also directly use the event name $E_1$ instead of the set transformer $[E_1]$, as well as $E$ instead of $[E]$.

In order to stepwise use patterns we need to show that each application of a pattern is correct, i.e.  that (\ref{eqn:refinement-scheduled}) holds. In order to do that, we assume that $\sched(S,X)$ represents a yet unscheduled loop of events $\sched(S,X)=(\e(S)\sqcap X)^{*};[\g(\e(S)\sqcap X)]$. We instantiate the existential quantifier in (\ref{eqn:decompostion-equality}) with $S_i$ as $E_i$. Hence,  we then need to show that $\{h\};\sched(E_1\rightarrow\{g\}\rightarrow S)=\{h\};X^{*};E_1;X^{*};\{g\};\sched(S)$.  Note that we also rely here on the properties (\ref{lem:preservation-1})-(\ref{lem:preservation-3}) in Lemma \ref{lem:preservation}. Note also that to ensure (\ref{eqn:refinement-scheduled}) we here assume $i\cap\neg \g(E \sqcap X)\subseteq \g(fin)$. The reason for formulating the pattern in this way is to be able to use the same verification approach also to nested loops.

\begin{lemma}{Context preservation.}\label{lem:preservation}
If $\{g\};S\sqsubseteq S;\{g\}$  then:
\begin{align}
&\{g\};S  =  \{g\};S;\{g\} \label{lem:preservation-1}\\
&\{g\};S^{*} = \{g\};S^{*};\{g\} \label{lem:preservation-2}\\
&\{g\};S^{*}  =  (\{g\};S)^{*} \label{lem:preservation-3}
\end{align}
\end{lemma}
\noindent The proofs of the properties in the lemma are straightforward and they are omitted for brevity. We can now prove the correctness of pattern $P_1$.
\begin{proof}
{\small
\[
\begin{array}{ll}
~&\{h\};\sched(E_1\rightarrow \{g\}\rightarrow S,X); [\neg \g(E_{1}\sqcap E\sqcap X)]\\
= & \{\mathrm{Representation~ of~}\sched(E_1\rightarrow \{g\}\rightarrow S) \}\\
~ & \{h\}; (E_{1}\sqcap E\sqcap X)^{{*}};[\neg \g(E_{1}\sqcap E\sqcap X)]\\
= & \{\mathrm{Decomposition~ \cite{Back-vonWright-99}:~}(S\sqcap T)^{*}=(S;T^{*})^{*};T^{*} \} \\
~ &  \{h\}; X^{{*}};(E_{1}\sqcap E;X^{{*}})^{{*}};[\neg \g(E_{1}\sqcap E\sqcap X)] \\
= & \{\mathrm{Distributivity}\}\\
~ &  \{h\}; X^{{*}}; ((E_{1};\, X^{{*}})\sqcap(E;\, X^{{*}}))^{{*}}; [\neg \g(E_{1}\sqcap E\sqcap X)]\\
= & \{\mathrm{Decomposition}\}\\
~ &  \{h\};  X^{{*}}; ((E_{1};\, X^{{*}})^{*};((E;\, X^{{*}});\,(E_{1};\, X^{{*}})^{{*}})^{{*}};[\neg \g(E_{1}\sqcap E\sqcap X)]\\
= &\{\mathrm{Unfolding~ (\ref{eqn:rule-unfolding})}\}\\
~ &  \{h\};  X^{{*}}; ((E_{1};\, X^{{*}}); (E_{1};\, X^{{*}})^{{*}})\sqcap \skipc;((E;\, X^{{*}});\,(E_{1};\, X^{{*}})^{{*}})^{{*}};[\neg \g(E_{1}\sqcap E\sqcap X)]\\
= &\{\mathrm{Assumption~3~ and~Property~ (\ref{lem:preservation-2})}\} \}\\
~ &  \{h\};  X^{{*}};  \{h\};(E_{1};\, X^{{*}}); (E_{1};\, X^{{*}})^{{*}})\sqcap \{h\};((E;\, X^{{*}});\,(E_{1};\, X^{{*}})^{{*}})^{{*}};[\neg \g(E_{1}\sqcap E\sqcap X)]\\
= &\{\mathrm{Distributivity, ~assumption~ } h\subseteq \neg \g(E)\mathrm{~and~ disabledness~of~guard}\}\\
~ &  \{h\};  X^{{*}};  \{h\};(E_{1};\, X^{{*}}); (E_{1};\, X^{{*}})^{{*}};((E;\, X^{{*}});\,(E_{1};\, X^{{*}})^{{*}})^{{*}};[\neg \g(E_{1}\sqcap E\sqcap X)]\\
= &\{\mathrm{Assumption}~ E_1=E_1;\{g\} \}\\
~ &  \{h\};  X^{{*}};  \{h\};E_{1};\, X^{{*}}; \{g\};(E_{1};\, X^{{*}})^{{*}};((E;\, X^{{*}});\,(E_{1};\, X^{{*}})^{{*}})^{{*}};[\neg \g(E_{1}\sqcap E\sqcap X)]\\
= & \{\mathrm{Assumption}~g\subseteq \neg \g(E_1)  \}\\
~ &  \{h\};  X^{{*}};  \{h\};E_{1};\, X^{{*}}; \{g\};(E;\, X^{{*}};(E_{1};\, X^{{*}})^{{*}})^{{*}};[\neg \g(E_{1}\sqcap E\sqcap X)]\\
\end{array}
\]
}
{\small
\[
\begin{array}{ll}
= & \{\mathrm{Property~ (\ref{lem:preservation-3})~ and~ *~ below}\}\\
~ &  \{h\};  X^{{*}};  \{h\};E_{1};\,\{g\}; X^{{*}}; \{g\};(\{g\};E;\, X^{{*}}; \{g\})^{{*}};[\neg \g(E_{1}\sqcap E\sqcap X)]\\
= & \{\mathrm{Leapfrog~ \cite{Back-vonWright-99}:}~ S;(T;S)^{*} = (S;T)^{*};S\}\\
~ &  \{h\};  X^{{*}};  \{h\};E_{1};\, \{g\};X^{{*}}; (\{g\};E;\, X^{{*}})^{{*}}; \{g\};[\neg \g(E_{1}\sqcap E\sqcap X)]\\
= & \{\mathrm{Assumption~} g\subseteq \neg \g(E_1) \mathrm{~ and~} \{g\};[g]=\{g\}\}\\
~ &  \{h\};  X^{{*}};  \{h\};E_{1};\,\{g\}; X^{{*}};  (\{g\};(E;\, X^{{*}}))^{{*}}; \{g\};[\neg \g (E\sqcap X)]\\
= & \{\mathrm{Lemma~9(c)~in~\cite{Back-vonWright-99}:}~ S^{*}=S^{*};S^{*} \mathrm{~and~ decomposition}\}\\
~ &  \{h\};  X^{{*}};  \{h\};E_{1};\{g\};X^{*};(\{g\};E \sqcap \{g\};X)^{*};[\neg \g (E\sqcap X)]\\
= & \{\mathrm{Property~(\ref{lem:preservation-2})~ and~assumption~ 5}\}\\
~ &  \{h\};  X^{{*}};  E_{1};X^{*};\{g\};(\{g\};E \sqcap \{g\};X)^{*};[\neg \g (E\sqcap X)]\\
= & \{\mathrm{Representation~ of~}\sched(S,X) \}\\
~& \{h\}; X^{{*}}; E_{1}; X^{{*}}; \{g\}; \sched(S,X)\\
\end{array}
\]
}
The proof of step $*$ is:
 {\small
 \[
 \begin{array}{ll}
 ~ & (\{\g\};E;\, X^{{*}};(E_{1};\, X^{{*}})^{{*}})^{{*}}\\
 = & \{\mathrm{Assumption~ 3~ and~ Properties~ (\ref{lem:preservation-1})~ and (\ref{lem:preservation-2})}\}\\
  ~ &(\{\g\};E;\, X^{{*}};\{\g\};(E_{1};\, X^{{*}})^{{*}})^{{*}}\\
  = & \{\mathrm{Assumption~2}\}\\
   ~ & (\{\g\};E;\, X^{{*}};\{\g\})^{{*}}\\
 \end{array}
 \]
 }
\end{proof}

\paragraph{Pattern 2}
The second pattern, $P_2$, also introduces sequential behaviour. However, this time we show that we can  group local behaviour $E_2$ to an arbitrary event. 
\begin{equation}\label{eqn:scheduling-pattern-1}
\begin{array}{l}
 P_{2}(E_1,E_2,h,g,S_1, X)\defs \\
 \begin{array}{lcl}
 \mathrm{Precondition} &: &h\\
 \mathrm{Schedule} &: &  E_1\rightarrow E_2\rightarrow \{g\}\rightarrow S  \\
 \mathrm{Result} &: & \{h\};X^{*};E_1;X^{*};E_2;X^{*};\{g\};\sched(S,X) \\
 \mathrm{Assumption~ 1} &: &  h\subseteq \neg \g(\e(S))  \\
 \mathrm{Assumption~ 2} &: & g\subseteq \neg \g(E_1\sqcap E_2)\\
 \mathrm{Assumption~ 3} &: & E_2;X= X;E_2\\
 \mathrm{Assumption~ 4} &: &\{\g(E_2)\};X= X;\{\g(E_2)\}\\
 \mathrm{Assumption~ 5} &: & \{g\};(X\sqcap \e(S))\sqsubseteq (X\sqcap \e(S));\{g\} \\
 \mathrm{Assumption~ 6} &: & \{h\};X\sqsubseteq X;\{h\} \\
 \mathrm{Assumption~7} & : & E_{2}=E_{2};\{g\}\\
\end{array}\\
\end{array}
\end{equation}
The assumptions in pattern $P_2$ are similar to the ones in $P_1$. However, we additionally need assumptions that states that $E_2$ and $X$ do not interfere with each other (assumptions 3 and 4). To prove the correctness of the pattern we need to show that
\begin{itemize}
\item By instantiation of (\ref{eqn:decompostion-equality}) we get: $\{h\};X^{*};E_1;X^{*};E_2;X^{*};\{g\};\sched(S,X)=\{h\};(E_1;E_2 \sqcap \e(S) \sqcap X)^{*};[\neg \g(E_1\sqcap E_2 \sqcap \e(S)\sqcap X)]$
\item Refinement (\ref{eqn:refinement-scheduled}): $\{h\};\sched(E_1\rightarrow E_2\rightarrow \{g\}\rightarrow S,X)\sqsubseteq\{h\};(E_1;E_2 \sqcap \e(S) \sqcap X)^{*};[\neg \g(E_1\sqcap E_2 \sqcap \e(S)\sqcap X)]$
\item Deadlock freedom: $\{h\};(E_1;E_2 \sqcap \e(S) \sqcap X)^{*};[\neg \g(E_1\sqcap E_2 \sqcap \e(S)\sqcap X)](\false)=\{h\};\sched(E_1\rightarrow E_2\rightarrow  \{g\} \rightarrow S,X)(\false)$
\end{itemize}
The deadlock freedom proof obligation ensures that the scheduling does not introduce new deadlocks. This was not needed in pattern $P_1$, as that pattern does not alter the behaviour of models. The proofs are straightforward using the assumptions in the pattern. This ensures that the scheduling does not introduce more deadlocks than in the original system.

\section{Scheduling of dining philosophers\label{sec:Dining-philosophers-2}}

We now return to the running example introduced in section \ref{sec:Dining-philosophers}. Up till now, the
dining philosophers model has been refined and split into sub-models. Now, we show how the sub-models can be turned
into tasks by introducing schedules. In the scheduling process we use the patterns given in section
\ref{sub:Introduction-of-schedules}. Correctness will be proven by checking the assumptions of
the patterns. We will concentrate on how to derive a schedule for task 1. The schedules for task
2, 3 and 4 can be derived analogously.

Our approach is that the schedule should be formulated such that it
fulfills the previously mentioned solution to the dining philosophers problem,
i.e., that each philosopher should pick up the lower numbered fork
first. Since we first want to pick up fork number 1, we wish
to schedule \emph{Ph1GetFork1} as the first event. The correct order of events
will be \emph{Ph1GetFork1}, \emph{Ph1GetFork2}, \emph{Ph1Eat}, \emph{Ph1RelFork2},
\emph{Ph1RelFork1}. This is captured by the following schedule:
\[
\begin{array}{l}\mathrm{Ph1GetFork1} \rightarrow \{g_1\} \rightarrow \mathrm{Ph1GetFork2} \rightarrow \mathrm{Ph1Eat} \rightarrow \{g_2\}\\
\rightarrow \mathrm{Ph1RelFork2} \rightarrow \{g_3\} \rightarrow \mathrm{Ph1RelFork1} \rightarrow \{g_4\}
\end{array}
\]
The assertions in the schedule are needed to capture intermediate results and thereby enable verification of the schedule in smaller parts.

We now want to prove that it is correct to schedule \emph{Ph1GetFork1} as
the first event. To show this, we will follow pattern $P_1$ introduced in
Section \ref{sub:Introduction-of-schedules} and show that the assumptions
1 - 5  for the pattern are fulfilled.
We instantiate pattern $P_1$ as $P_1(Ph1GetFork1, h_1, g_1, S_r, X_{t1})$, where
$h_1 = (fork1 \neq 1 \wedge ph1eaten = FALSE)$,
$g_1 = (fork1 = 1 \vee ph1eaten = TRUE)$, $S_r = \mathrm{Ph1GetFork2} \rightarrow
\mathrm{Ph1Eat} \rightarrow \{g_2\} \rightarrow \mathrm{Ph1RelFork2} \rightarrow
\{g_3\} \rightarrow \mathrm{Ph1RelFork1} \rightarrow \{g_4\}$
and $X_{t1} = $ $\{\mathrm{Ph2GetFork2}$, $\mathrm{Ph4GetFork1}$, $\mathrm{Ph2RelFork2}$,
$\mathrm{Ph4RelFork1}\}$. 

We chose precondition $h_1$ so that it also is an invariant for the external
events $X_{t1}$. Here, $h_1$ states that philosopher 1 does not hold his
forks nor has he eaten. Moreover, we chose assertion $g_1$ to state that philosopher 1
has picked up fork 1 or eaten. This condition is an invariant for the events
$\e(S_r) \cup X_{t1}$ and established by \emph{Ph1GetFork1}. We now confirm that
the assumptions for the pattern hold:

\begin{itemize}
\item 
$h_1 = (fork1 \neq 1 \wedge ph1eaten = FALSE)$ implies that events in $\e(S_r)$ are disabled. This holds, since they are only enabled when philosopher 1 holds fork 1 or has eaten.
\item 
The assertion $g_1 = (fork1 = 1 \vee ph1eaten = TRUE)$ following event \emph{Ph1GetFork1} ensures that \emph{Ph1GetFork1} is disabled. Since $g_1$ is a negation of the guard of \emph{Ph1GetFork1} the second assumption is fulfilled.
\item 
$g_1$ is an invariant of the environment $\e(S_r) \cup X_{t1}$. This is fulfilled, since in the events of $\e(S_r)$ philosopher 1 holds fork 1 or has eaten. Moreover, the events in $X_{t1}$ that share fork 1 are not enabled when philosopher 1 holds fork 1, and none of these events modify variable $ph1eaten$.
\item 
$h_1$ is an invariant of the external events $X_{t1}$. Since none of the external events model that philosopher 1 picks up fork 1 or modify variable $ph1eaten$, this assumption holds.
\item 
Event \emph{Ph1GetFork1} establishes $g_1$. This holds trivially since \emph{Ph1GetFork1} models that philosopher 1 picks up fork 1 ($fork1:=1$).
\end{itemize}

To verify the complete schedule, we then apply pattern $P_2$ once, followed by three applications of $P_1$. In the last application of $P_1$, the schedule following the assertion is empty. This can be interpreted as a schedule with an event that is always disabled. When task 1 has been fully proven, the whole procedure is repeated to schedule tasks 2, 3 and 4 in the order shown in the table below (for simplicity, the assertions are not shown).\\

\begin{tabular}{|l|l|l|l|}
\hline 
\hspace{5mm}Task 1 & \hspace{5mm}Task 2 & \hspace{5mm}Task 3 & \hspace{5mm}Task 4\tabularnewline
\hline 
\hspace{3.5mm} Ph1GetFork1 & \hspace{3.5mm} Ph2GetFork2 & \hspace{3.5mm} Ph3GetFork3 & \hspace{3.5mm} Ph4GetFork1\tabularnewline
 $\rightarrow$ Ph1GetFork2\hspace{2mm} & $\rightarrow$ Ph2GetFork3\hspace{2mm} &  $\rightarrow$ Ph3GetFork4\hspace{2mm} & $\rightarrow$ Ph4GetFork4\hspace{2mm}\tabularnewline
$\rightarrow$ Ph1Eat & $\rightarrow$ Ph2Eat & $\rightarrow$ Ph3Eat & $\rightarrow$ Ph4Eat\tabularnewline
$\rightarrow$ Ph1RelFork2 & $\rightarrow$ Ph2RelFork3 & $\rightarrow$ Ph3RelFork4 & $\rightarrow$ Ph4RelFork4\tabularnewline
$\rightarrow$ Ph1RelFork1 & $\rightarrow$ Ph2RelFork2 & $\rightarrow$ Ph3RelFork3 & $\rightarrow$ Ph4RelFork1\tabularnewline
\hline
\end{tabular}

\section{Conclusions and related work\label{sec:Conclusions-and-related}}

In this paper, we have proposed a method of correct-by construction development of concurrent programs using Event-B. The programs are first developed as proposed by Hoang and Abrial \cite{Hoang-Abrial-10}. From this development process we obtain a number of sub-models  that communicate via shared variables, which represent the program. We then introduce explicit control flow in the form of schedules  for each  sub-model, so that each sub-model/schedule corresponds to exactly one task. The schedules are introduced as correctness preserving refinements. We use a set-transformer semantics for Event-B, as well as well known algebraic rules \cite{Back-vonWright-99} for the analysis of correctness. The schedules are verified in a stepwise manner, and each step carries some related proof obligations. The schedules enable more efficient implementation of the Event-B models as more explicit control flow information is available than for pure event-B models. We can, e.g., use the transformations in \cite{Bostrom-10} to introduce traditional control flow constructs, such as while loops and if-statements, as well as remove unnecessary guards. Furthermore, the schedules give a process-oriented specification of the behaviour of the models.

Our goal is to compositionally reason about concurrent programs. This has been a very active field of research \cite{deRoever-01}. Our approach directly extends the approach in \cite{Hoang-Abrial-10} for development of concurrent programs with explicit schedules of events.  Compositional reasoning in this setting goes back to the work of Owicki and Gries \cite{Owicki-Gries-76} and Jones' Rely-Guarantee reasoning \cite{Jones-83}. The decomposition method based on shared variables in Event-B \cite{Abrial-Hallerstede-07,Hoang-Abrial-10} is based on these ideas. Essentially the same approach is also available for action systems using the refinement calculus  \cite{BaWr03}. The theory for decomposition in the  set-transformer setting is largely based on that paper.   Several approaches to introducing control flow into Event-B models have been developed. Hallerstede's approach in \cite{Hallerstede-10} to adding control flow only deals with sequential programs and it is thus more related to Bostr\"om's earlier work \cite{Bostrom-10}. The scheduling approaches in \cite{Ilisov-09,Schneider-Treharne-Wehrheim-10} can also handle concurrent schedules. In \cite{Ilisov-09} the scheduling (referred to as {\em flows}) is expressed using a special purpose language, while in the approach \cite{Schneider-Treharne-Wehrheim-10} the scheduling is expressed in CSP. The latter approach can be seen as an extension of the former. Processes or flows are both considered to communicate via shared events.    Our focus is on compositional verification and scheduling of concurrent programs that use shared variables for communication. However, in both approaches not all events need to be scheduled, but non-scheduled events are considered interleaved in the scheduled. This could be used to take into account external events, and thus be used for compositional verification of shared variable programs also. Our contribution is threefold: 1) Compared to purely event-based modelling, we consider explicit schedules of events that can be interleaved 2) We do all analysis on the level of set transformers, which gives  convenient formalism to algebraically perform the needed analysis of Event-B models 3) We provide patterns and a method to develop patterns for introducing control flow in a stepwise manner. This is important, since verifying that a certain event schedule is correct can be very challenging and reusable scheduling structures can significantly aid in this task.

Set-transformers give a powerful framework to reason about Event-B models on a high level of abstraction. They give a good basis for creating reusable patterns for scheduling, which are essential for practical applications. If schedules are introduced as a last refinement step, as in the example of this paper, existing tool support can be used for development up till, but not including, the scheduling step. Future work involves investigating tool support for schedule application. Generation of refinement proof obligations for scheduled models is also of interest, since that would allow for schedule intoduction earlier in the refinement chain.

\bibliographystyle{eptcs}
\bibliography{sched}

\end{document}